\documentclass[preprint,12pt, a4paper]{elsarticle}
\biboptions{sort&compress}

\usepackage{amssymb}
\usepackage{hyperref}
\usepackage{amsmath}
\usepackage{indentfirst}
\usepackage{array}
\usepackage{booktabs}
\usepackage{xltabular}
\usepackage{caption}
\usepackage{float}

\usepackage{xcolor}
\usepackage{listings}

\let\origthelstnumber\thelstnumber

\makeatletter

\newcommand*\Suppressnumber{%
    \lst@AddToHook{OnNewLine}{%
        \let\thelstnumber\relax
        \advance\c@lstnumber-\@ne\relax
    }%
}

\newcommand*\Reactivatenumber{%
    \lst@AddToHook{OnNewLine}{%
        \let\thelstnumber\origthelstnumber
        \advance\c@lstnumber\@ne\relax
    }%
}

\makeatother

\definecolor{codebackground}{HTML}{F7F9FC}
\definecolor{codeframe}{HTML}{CCD5E0}
\definecolor{codekeyword}{HTML}{174A8B}
\definecolor{codecomment}{HTML}{2F6F44}
\definecolor{codestring}{HTML}{8A4B20}
\definecolor{codenumber}{HTML}{7A8491}
\definecolor{codetext}{HTML}{20252B}

\lstdefinestyle{algoplasma}{
    language=Fortran,
    backgroundcolor=\color{codebackground},
    basicstyle=\ttfamily\fontsize{7.7}{9.2}\selectfont\color{codetext},
    keywordstyle=\bfseries\color{codekeyword},
    commentstyle=\ttfamily\color{codecomment},
    stringstyle=\ttfamily\color{codestring},
    numbers=left,
    numberstyle=\ttfamily\tiny\color{codenumber},
    numbersep=9pt,
    stepnumber=1,
    frame=single,
    framerule=0.35pt,
    rulecolor=\color{codeframe},
    framesep=6pt,
    xleftmargin=2.6em,
    framexleftmargin=2.1em,
    xrightmargin=0.4em,
    framexrightmargin=0.4em,
    aboveskip=0.9em,
    belowskip=0.7em,
    captionpos=b,
    columns=fullflexible,
    keepspaces=true,
    breaklines=true,
    breakatwhitespace=true,
    breakautoindent=true,
    breakindent=1.5em,
    showstringspaces=false,
    tabsize=4,
    escapeinside={(*@}{@*)}
}

\journal{SoftwareX}

\begin{document}
\renewcommand{\labelenumii}{\arabic{enumi}.\arabic{enumii}}

\begin{frontmatter}



\title{AlgoPlasma: Open Algorithms for Plasma Modeling}


\author[hit]{Yinjian Zhao\corref{cor1}}
\ead{zhaoyinjian@hit.edu.cn}
\ead[url]{https://homepage.hit.edu.cn/zhaoyinjian}
\cortext[cor1]{Corresponding author.}
\author[hit]{Zhongping Zhao}
\author[hit]{Zhe Liu}
\author[hit]{Baisheng Wang}
\author[hit]{Zilong Peng}
\author[hit]{Xin Luo}
\author[hit]{Lihuan Xie}
\author[hit]{Xi Chen}
\author[hit]{Zhijun Zhou}
\author[hit]{Kunpeng Zhong}
\author[hit]{Yingjie Chen}
\author[hit]{Changzheng Hu}

\address[hit]{School of Energy Science and Engineering, Harbin Institute of Technology, Harbin 150001, People's Republic of China}

\begin{abstract}

AlgoPlasma is an open-source library in which core numerical algorithms
for plasma modeling are implemented as modular, well-documented, and
independently testable components.
Rather than offering a complete simulation code, it allows researchers to select, adapt, and assemble the required components into application-specific workflows.
The current release is centered on particle-based simulation, while AlgoPlasma is designed to encompass a broader range of approaches to plasma modeling.
It provides components for particle initialization and advancement,
particle--grid coupling, field solution, collision modeling,
neutral transport, selected continuity updates, probe signal
analysis, and input/output, together with Message Passing
Interface (MPI) exchange routines and selected Open
Multi-Processing (OpenMP) kernels.
Documentation links mathematical
formulations to source implementations, interfaces, and usage, while
verification and validation cases evaluate numerical accuracy and physical
behavior. 
AlgoPlasma thus establishes a shared algorithmic foundation for plasma modeling, transforming repeatedly reimplemented numerical methods into open, reusable, tested, and explainable components for research, verification, education, and collaborative development.

\end{abstract}


\end{frontmatter}


\begin{table}[htbp!]
\centering
\caption{Code metadata (mandatory)}
\label{tab:code_metadata}

\begingroup
\footnotesize
\setlength{\tabcolsep}{4pt}
\renewcommand{\arraystretch}{1.1}

\begin{tabularx}{\linewidth}{
|l|
>{\raggedright\arraybackslash}p{0.38\linewidth}|
>{\raggedright\arraybackslash}X|
}
\hline
\textbf{Nr.} &
\textbf{Code metadata description} &
\textbf{Metadata} \\
\hline

C1 & Current code version & \texttt{v1.0.1} \\
\hline

C2 & Permanent GitHub link to code/repository used for this code version &
Release: \url{https://github.com/AlgoPlasma/AlgoPlasma/releases/tag/v1.0.1}
\newline
Commit SHA: \path{adff2dcfb78fb39bb939cc9db80d63c6ccfe798c}
\newline
Archival DOI: \url{https://doi.org/10.5281/zenodo.23227612} \\
\hline

C3 & Legal Code License & Apache License 2.0 (Apache-2.0) \\
\hline

C4 & Code versioning system used & Git \\
\hline

C5 & Software code languages, tools, and services used &
Fortran, C/C++, Python 3, Bash; Doxygen, Graphviz; Read the Docs \\
\hline

C6 & Compilation requirements, operating environments \& dependencies &
Linux (developed on Ubuntu 22.04 LTS).
\newline
Build: GNU Fortran, GCC/G++, CMake, Make.
\newline
Component specific: HYPRE$\geq$3.0 (Poisson solvers),
MPI (MPI-dependent routines), HDF5 (HDF5 I/O),
OpenMP (parallel kernels);
C++20 (\texttt{G02});
CMake$\geq$3.20 (\texttt{G02} and the two-stream example);
Python$\geq$3.10 (\texttt{K04}), NumPy (\texttt{K01}--\texttt{K04}).
\newline
Postprocessing: NumPy, SciPy, Matplotlib, ImageIO.
\newline
Documentation: Sphinx, Breathe, Doxygen, Graphviz. \\
\hline

C7 & If available Link to developer documentation/manual &
\url{https://algoplasma.readthedocs.io/en/v1.0.1/} \\
\hline

C8 & Support email for questions &
\href{mailto:contact@algoplasma.com}{contact@algoplasma.com} \\
\hline
\end{tabularx}

\endgroup
\end{table}

\section{Motivation and significance}

Numerical simulation is an essential tool for investigating plasma behavior across a wide range of scientific and engineering applications, such as laboratory and space plasmas, fusion energy,
plasma-based acceleration, and electric propulsion, among others. Plasma dynamics involve nonlinear, collective, and multiscale processes and are therefore simulated using fluid, kinetic, hybrid, or reduced models according to the physical regime and research objective. Among particle-based kinetic approaches, the particle-in-cell (PIC) method self-consistently evolves particle distributions and electromagnetic fields and has become central to computational plasma research \cite{birdsall1991,hockney1988,verboncoeur2005}.

Despite differences in their physical descriptions, these models rely on recurring numerical operations, including 
field solution, collision treatment, continuum updates, parallel data
exchange, and diagnostic input/output; particle-based and hybrid methods
additionally require particle advancement and particle--grid transfer.
Although the underlying methods are well established, their implementation requires careful treatment of time integration, coordinate systems, boundary conditions, stability, conservation, and computational efficiency. Reimplementing these operations for each application duplicates development effort and may leave important numerical assumptions implicit. Conversely, implementations embedded in complete simulation codes are often coupled to application-specific data structures and execution procedures, which can complicate the independent inspection, testing,
modification, and reuse of individual algorithms.
Established PIC codes such as WarpX, OSIRIS, and EPOCH exemplify this
integrated approach, providing comprehensive simulation capabilities
through large and complex codebases
\cite{vay2018warpx,osiris,arber2015epoch}.

In response to these needs, a range of software resources supports
numerical implementation and application development at different
levels.
UPIC provides PIC component libraries and reference applications,
while PICSAR supplies optimized numerical kernels \cite{upic,picsar}.
UCLA PIC Skeleton Codes offer compact educational and benchmark
implementations with reusable procedures \cite{skeleton}.
PICMI defines a common interface for PIC simulation configuration
\cite{picmi}, whereas AMReX provides mesh and particle infrastructure
for application development \cite{amrex}.
Tab.~\ref{tab:software_comparison} compares these representative
resources in terms of their numerical scope, interfaces,
implementation requirements, and external reuse.
Within this ecosystem, AlgoPlasma provides a collection of standalone implementations of plasma modeling algorithms as inspectable, testable, adaptable, and reusable source components, without prescribing a complete simulation workflow.

\begin{table}[htbp!]
\centering
\caption{Comparison of AlgoPlasma with representative software
resources: (A) numerical scope and interfaces.}
\label{tab:software_comparison}

\begingroup
\scriptsize
\setlength{\tabcolsep}{3pt}
\renewcommand{\arraystretch}{1.1}

\begin{tabularx}{\linewidth}{
@{}
>{\hsize=0.857143\hsize\linewidth=\hsize
  \raggedright\arraybackslash}X
>{\hsize=1.071429\hsize\linewidth=\hsize
  \raggedright\arraybackslash}X
>{\hsize=1.017857\hsize\linewidth=\hsize
  \raggedright\arraybackslash}X
>{\hsize=0.964286\hsize\linewidth=\hsize
  \raggedright\arraybackslash}X
>{\hsize=1.392857\hsize\linewidth=\hsize
  \raggedright\arraybackslash}X
>{\hsize=0.696428\hsize\linewidth=\hsize
  \raggedright\arraybackslash}X
@{}
}
\toprule
\textbf{Software} &
\textbf{Scope} &
\textbf{Interface granularity} &
\textbf{Coordinate systems} &
\textbf{Representative algorithms} &
\textbf{Language} \\
\midrule

AlgoPlasma &
Plasma modeling source components &
Subroutines and modules &
Cartesian; cylindrical; selected nonuniform grids &
Pushers; scatter/\allowbreak{}gather;
Poisson/\allowbreak{}Maxwell; MCC;
neutral transport; selected continuity updates;
probe signal analysis &
Fortran; C/\allowbreak{}C++; Python \\
\addlinespace[2pt]

UPIC \cite{upic} &
PIC component libraries and reference applications &
Procedure libraries and module interfaces &
Cartesian, 1D--3D &
Electrostatic, electromagnetic and Darwin PIC;
spectral field solvers &
Fortran; Python drivers \\
\addlinespace[2pt]

PICSAR \cite{picsar} &
Optimized PIC kernels and quantum electrodynamics (QED) routines &
Fortran kernels; Python wrapper; C++ QED headers &
Mainly Cartesian, 2D/\allowbreak{}3D &
Boris/\allowbreak{}Vay pushers;
direct/\allowbreak{}Esirkepov current deposition;
finite difference/\allowbreak{}spectral Maxwell solvers;
QED processes &
Fortran; C++; Python \\
\addlinespace[2pt]

UCLA PIC Skeleton Codes \cite{skeleton} &
Educational and benchmark PIC codes &
Compact drivers and array based procedure libraries &
Cartesian, 1D--3D &
Electrostatic, electromagnetic and Darwin PIC;
FFT field solvers &
Fortran; C; selected Python/\allowbreak{}CUDA \\
\addlinespace[2pt]

PICMI \cite{picmi} &
PIC configuration standard &
Standardized configuration classes &
Cartesian/\allowbreak{}cylindrical grid specifications &
Grid, species, solver and diagnostic configuration;
no PIC kernels &
Python reference implementation \\
\addlinespace[2pt]

AMReX \cite{amrex} &
Block structured adaptive mesh refinement (AMR)
and particle infrastructure &
Mesh/\allowbreak{}particle containers and solver classes &
Cartesian, 1D--3D;
selected 2D cylindrical/\allowbreak{}1D spherical operators &
AMR; particle operations; linear solvers;
load balancing &
C++; optional Fortran interfaces \\

\bottomrule
\end{tabularx}

\endgroup
\end{table}

\begin{table}[htbp!]
\centering
\begingroup

\addtocounter{table}{-1}
\ifdefined\theHtable
\renewcommand{\theHtable}{\arabic{table}.continued}
\fi

\caption{Comparison of AlgoPlasma with representative software
resources (continued): (B) implementation and external reuse.}
\label{tab:software_comparison_b}

\scriptsize
\setlength{\tabcolsep}{3pt}
\renewcommand{\arraystretch}{1.1}

\begin{tabularx}{\linewidth}{
@{}
>{\hsize=0.857143\hsize\linewidth=\hsize
  \raggedright\arraybackslash}X
>{\hsize=1.232143\hsize\linewidth=\hsize
  \raggedright\arraybackslash}X
>{\hsize=1.178571\hsize\linewidth=\hsize
  \raggedright\arraybackslash}X
>{\hsize=0.857143\hsize\linewidth=\hsize
  \raggedright\arraybackslash}X
>{\hsize=0.803571\hsize\linewidth=\hsize
  \raggedright\arraybackslash}X
>{\hsize=1.071429\hsize\linewidth=\hsize
  \raggedright\arraybackslash}X
@{}
}
\toprule
\textbf{Software} &
\textbf{Main dependencies} &
\textbf{Verification evidence} &
\textbf{Parallel capability} &
\textbf{License} &
\textbf{External reuse} \\
\midrule

AlgoPlasma &
MPI/\allowbreak{}OpenMP/\allowbreak{}HYPRE/\allowbreak{}HDF5/\allowbreak{}NumPy
by component &
Maxwell MMS and component tests;
varying coverage &
MPI exchange; selected OpenMP kernels &
Apache-2.0 &
Compile or import selected sources \\
\addlinespace[2pt]

UPIC \cite{upic} &
Make; MPI/\allowbreak{}OpenMP; optional Python &
Reference applications and selected field tests &
MPI/\allowbreak{}OpenMP; vectorization &
Custom commercial-reservation license &
Build libraries; call from own driver \\
\addlinespace[2pt]

PICSAR \cite{picsar} &
MPI/\allowbreak{}OpenMP;
FFTW/\allowbreak{}Forthon by configuration;
Boost for QED table generation &
PIC unit/\allowbreak{}acceptance tests;
separate QED tests &
MPI/\allowbreak{}OpenMP; vectorization;
GPU compatible QED kernels &
BSD-3-Clause-LBNL &
Link library; use Python wrapper or include QED headers \\
\addlinespace[2pt]

UCLA PIC Skeleton Codes \cite{skeleton} &
Make;
MPI/\allowbreak{}OpenMP/\allowbreak{}CUDA/\allowbreak{}Python
by example &
Reference outputs and energy diagnostics &
MPI/\allowbreak{}OpenMP/\allowbreak{}CUDA variants;
vectorization &
BSD-2-Clause in inspected examples &
Build examples; reuse procedures \\
\addlinespace[2pt]

PICMI \cite{picmi} &
Python; NumPy/\allowbreak{}SciPy;
compatible backend for execution &
Interface/\allowbreak{}example tests;
numerical verification by backend &
Backend dependent &
BSD-3-Clause-LBNL &
Use backend PICMI classes \\
\addlinespace[2pt]

AMReX \cite{amrex} &
CMake/\allowbreak{}Make;
optional MPI/\allowbreak{}OpenMP and GPU toolchains &
CTest and nightly regressions &
MPI/\allowbreak{}OpenMP;
CUDA/\allowbreak{}HIP/\allowbreak{}SYCL &
BSD-3-Clause &
Build/\allowbreak{}link library; use framework APIs \\

\bottomrule
\end{tabularx}

\endgroup
\end{table}

AlgoPlasma, formerly known as the Plasma Modeling Subprogram Library
(PMSL), originated as an in-house algorithm collection for
electric propulsion research. Its components are organized by
numerical task, with the current release covering particle
initialization and pushing, particle--grid coupling, Poisson and
Maxwell solvers, collision models, neutral transport, selected
continuity updates, probe signal analysis, parallel data exchange,
and input/output. Support for Cartesian and cylindrical geometries
depends on the selected component.
Researchers can use the interface documentation and available
test cases to select, adapt, and integrate components into
application-specific programs while retaining control over
the physical model and simulation procedure.
Implemented primarily in Fortran, with C interfaces to HYPRE,
a C++ collision component, and Python routines for diagnostics,
initialization, and postprocessing, AlgoPlasma provides Message
Passing Interface (MPI) exchange routines and selected
computational kernels parallelized with Open Multi-Processing
(OpenMP).
The current release does not provide general-purpose C/C++ or
Python bindings to the Fortran components.

Although artificial intelligence (AI) tools have made code generation
and cross-language translation increasingly accessible
\cite{chen2021codex,roziere2020transcoder}, a central challenge remains:
translating a numerical method from the literature into a modular
implementation whose accuracy and numerical behavior have been
rigorously verified.
AlgoPlasma addresses this challenge through documented
implementations and available test cases.
Users can progress from understanding a method
to running examples and tests, and ultimately to adapting
and integrating individual components into a coupled model,
without first having to master the architecture of a complete
plasma application.

The remainder of this article is organized as follows.
Sec.~\ref{arch} describes the architecture and core functionality
of AlgoPlasma. Sec.~\ref{example} presents the two-stream PIC
example and its numerical validation. Sec.~\ref{impact} discusses
scientific impact, maintenance and development, and current
limitations. Sec.~\ref{conclusion} concludes the article.

\section{Software description}
\label{arch}

\subsection{Library architecture}

\begin{figure}[htbp]
    \centering
    \includegraphics[width=\linewidth]{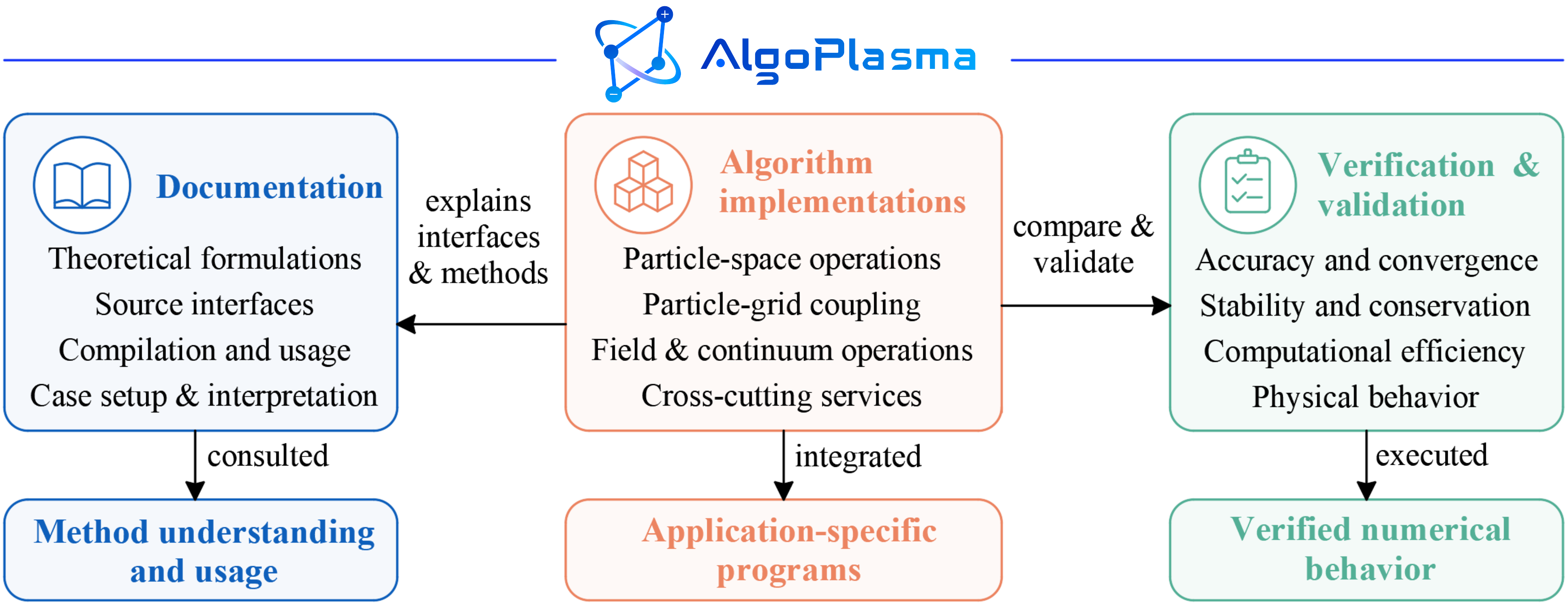}
    \caption{Overall architecture of AlgoPlasma as a reusable algorithm library.}
    \label{fig:library_architecture}
\end{figure}

AlgoPlasma integrates algorithm implementations,
documentation, and focused verification and validation (V\&V)
cases within the library architecture shown in
Fig.~\ref{fig:library_architecture}.

Organized by computational task, the numerical core comprises
source components that are selectively compiled, linked, or
imported into applications, without requiring prior compilation
and installation of a unified binary library.
The modules operate on caller-managed data, while the application
driver controls data organization, physical coupling, boundary
configuration, and execution order.
The demonstrated component combinations require no modifications
to component source files; integration is handled by the
application driver and any necessary adapters.
Supporting utilities provide selected solver interfaces, data
preparation, analysis, and visualization, with external dependencies
introduced only where required.

Grid-based components use structured Cartesian or cylindrical
meshes, with selected implementations supporting nonuniform
cylindrical meshes.
Applications supply the required grid spacings or coordinate
arrays and arrange particle and field arrays according to each
routine's prescribed shapes, component ordering, index bounds,
field locations, and guard layers.
Compatible externally generated meshes can therefore be used
after application-side conversion to these representations,
without requiring a common mesh object; general-purpose mesh
import and unstructured-grid operators are not provided.
Both dimensional and normalized calculations are supported,
provided that inputs and coefficients are consistent with the
unit conventions of the selected routines.

Most core components are accompanied by focused V\&V cases that
directly call the routines used in applications.
These cases assess numerical accuracy, convergence, stability,
conservation, or expected physical behavior against analytical
solutions, reference results, or established benchmarks.
The documentation records interfaces, usage, case configurations,
and interpretation of results.
Narrative pages are generated with Sphinx; application programming
interface (API) information is extracted with Doxygen and
integrated through Breathe.

Original AlgoPlasma source code is distributed under the Apache
License 2.0. License headers are not yet uniform across
the source tree. Third-party dependencies and any redistributed
third-party code or data remain subject to their respective licenses.

\subsection{Core functionality}

\begin{figure}[htb]
    \centering
    \includegraphics[width=\linewidth]{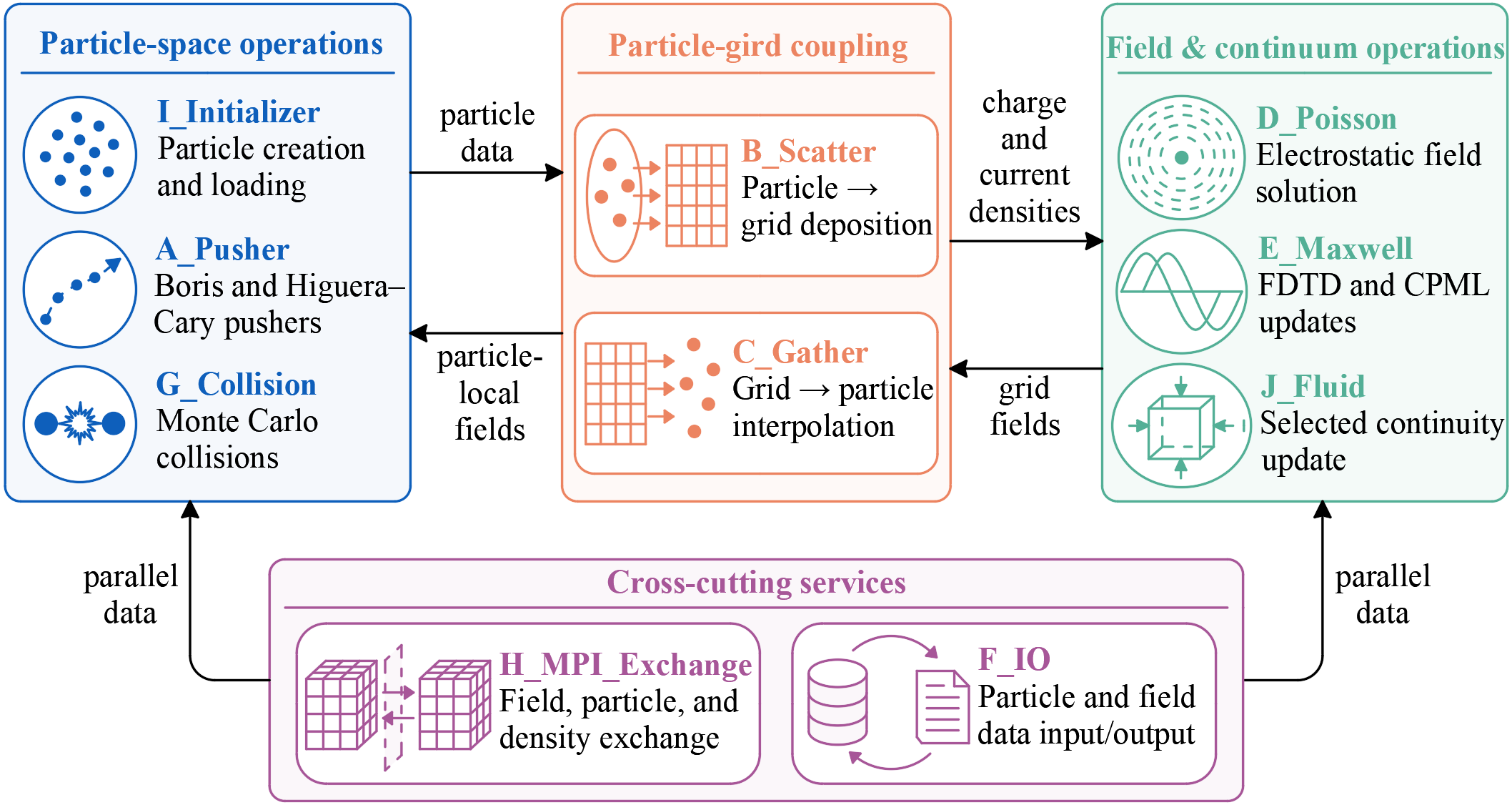}
    \caption{Functional organization of AlgoPlasma simulation components
    and their principal data exchanges.}
    \label{fig:functional_organization}
\end{figure}

The current functionality of AlgoPlasma is organized into eleven
letter-coded categories, \texttt{A}--\texttt{K}.
Each category corresponds to a distinct computational function,
while alternative algorithmic implementations within a category
are distinguished by numerical identifiers, such as
\texttt{A01} and \texttt{A02}.
The distributed grid-based implementations are primarily 3D,
with axisymmetric 2D Maxwell updates in \texttt{E01}
and selected 2D $r$--$z$ neutral transport and continuity
routines in \texttt{J};
no dedicated 1D spatial solvers are provided.
\ref{app:component_inventory} summarizes each component's
numerical task, method, geometry, dimensionality, and test availability.
For clarity, categories \texttt{A}--\texttt{J} are grouped
according to the data on which they operate and their roles
in plasma simulation, as shown in
Fig.~\ref{fig:functional_organization}.
The connections in the figure represent typical data exchanges
rather than a prescribed execution sequence.

\begin{itemize}

\item \textbf{Particle-space operations.}
These components operate directly on particle phase-space data. \texttt{I\_Initializer} provides particle creation and loading procedures, while \texttt{A\_Pusher} advances particle positions and velocities using algorithms including the Boris and Higuera--Cary pushers
\cite{boris1970,higuera2017}. 
\texttt{G\_Collision} implements Monte Carlo collision (MCC)
models using null-collision sampling \cite{vahedi1995},
with a C++ component for data-driven reaction networks.

\item \textbf{Particle--grid coupling.}
These components provide bidirectional transfer between particle and grid representations. \texttt{B\_Scatter} deposits particle quantities, such as charge and current densities, onto the computational grid, whereas \texttt{C\_Gather} interpolates grid-based fields to particle positions.
They therefore connect particle-space operations with mesh-based field
calculations \cite{birdsall1991,hockney1988}.

\item \textbf{Field and continuum operations.}
These components operate primarily on grid-based quantities. \texttt{D\_Poisson} provides Poisson solvers and electric field
evaluation for electrostatic calculations, including interfaces to the
High Performance Preconditioners (HYPRE) library
\cite{falgout2002hypre}.
\texttt{E\_Maxwell} implements finite-difference time-domain (FDTD)
updates and convolutional perfectly matched layer (CPML) boundary
treatments \cite{yee1966,roden2000cpml}.
\texttt{J\_Fluid} provides reduced $r$--$z$ neutral transport
with two velocity components, using free molecular Monte Carlo
and discrete ordinates, together with continuity updates based
on prescribed velocities or reference face fluxes.

\item \textbf{Cross-cutting services.}
These components support multiple numerical modules rather than representing a specific physical process. \texttt{H\_MPI\_Exchange} handles the exchange of field, particle, and density data between subdomains in distributed-memory calculations, while \texttt{F\_IO} provides input and output routines for particle and field data.

\end{itemize}

The components can be assembled at different levels of complexity. A single-particle study may require only initialization and particle pushing, whereas an electrostatic PIC application may combine particle pushing, deposition, interpolation, and Poisson solution. Electromagnetic or distributed-memory applications can additionally incorporate Maxwell updates, MPI data exchange, and input/output as needed.
Note that boundary operations are incorporated into the relevant components.
During assembly, the application configures the selected
operations, supplies the required data, and implements
any additional physical boundary models.
Tab.~\ref{tab:boundary_treatment} summarizes this division of
responsibility for the principal module families.

\begin{table}[htbp]
\centering
\caption{Boundary operations and application responsibilities.}
\label{tab:boundary_treatment}

\begingroup
\scriptsize
\setlength{\tabcolsep}{4pt}
\renewcommand{\arraystretch}{1.15}

\begin{tabularx}{\linewidth}{
@{}
>{\raggedright\arraybackslash}p{0.13\linewidth}
>{\raggedright\arraybackslash}X
>{\raggedright\arraybackslash}X
@{}
}
\toprule
\textbf{Module family} &
\textbf{Available boundary operations} &
\textbf{Application responsibilities} \\
\midrule

Poisson &
\texttt{D01}: boundary assembly and periodic settings through
case-specific include files.
\newline
\texttt{D02}--\texttt{D04}: Dirichlet, Neumann, and periodic
boundaries; surface charge and Robin-type outflow corrections.
\newline
\texttt{D03}/\texttt{D04}: cylindrical axis closure.
\newline
\texttt{D05}: periodic potential guard cell filling. &
Select boundary types and supply boundary data, including surface
charge; set consistent nonperiodic potential guard values for
field reconstruction. \\

\addlinespace[4pt]

Maxwell &
\texttt{E01}--\texttt{E03}: CPML absorbing updates.
\newline
\texttt{E01}/\texttt{E02}: cylindrical axis closure. &
Specify CPML coefficients and update regions; maintain memory
arrays; implement other boundary conditions and guard cell
updates. \\

\addlinespace[4pt]

Fluid &
\texttt{J01}/\texttt{J02}: inflow, open outflow,
and specular/diffuse reflection.
\newline
\texttt{J03}: prescribed inflow, reference-based outflow,
and impermeable walls.
\newline
\texttt{J01} continuity utilities: fluxes from exterior
states or prescribed boundary fluxes. &
Specify inlet and wall conditions; supply reference density,
face fluxes, and exterior states as required. \\

\addlinespace[4pt]

Particle-related &
\texttt{B02}: azimuthal periodic indexing and axis averaging.
\newline
\texttt{C03}: optional angular wrapping for interpolation.
\newline
\texttt{H02}: periodic particle wrapping and removal at
nonperiodic outer boundaries. &
Handle deposition boundaries and interpolation guard fields;
implement reflection, injection/emission, and wall charging
as needed. \\

\bottomrule
\end{tabularx}

\par\smallskip
\begin{minipage}{\linewidth}
\footnotesize
\textit{Notes.}
\texttt{D03}/\texttt{D04} support periodicity only in the
azimuthal and axial directions.
\texttt{H02} particle removal does not calculate wall charge
or energy deposition.
\end{minipage}

\endgroup
\end{table}

In addition to these simulation components,
\texttt{K\_Diagnostics} provides NumPy based Python routines
for experimental or simulated probe time series.
Its components perform spectral and cross-phase analysis,
estimate wavenumber--frequency spectra along a probe baseline,
infer two-dimensional wavevectors, and extract phase-averaged
breathing waveforms from repeated records at a single point.

\section{Illustrative example}
\label{example}

To illustrate how AlgoPlasma components can be assembled into an
application-specific program, we consider the electrostatic two-stream
instability in two spatial and three velocity dimensions (2D3V)
\cite{birdsall1991,hockney1988,morse1969multid}.

\subsection{Physical and numerical setup}

The system consists of two equally populated Maxwellian electron
beams with drift velocities $\pm3v_{te}$ along the $x$ direction
and a uniform, immobile neutralizing background.
The normalization adopts
$v_{te}=\sqrt{k_{\mathrm{B}}T_e/m_e}$ and
$\omega_{pe}=\sqrt{n_0e^2/(m_e\varepsilon_0)}$
as fundamental reference scales, with the associated length scale given
by the Debye length $\lambda_{De}=v_{te}/\omega_{pe}$, where $n_0$ is the
initial total electron density.
The dimensionless variables used in the code are summarized in
Tab.~\ref{tab:two_stream_normalization}.

The periodic domain,
$64\lambda_{De}\times64\lambda_{De}$,
is discretized on a $64\times64$ Cartesian grid.
Each beam is represented by 64 macroparticles per cell,
giving 524,288 electron macroparticles in total.
Particles are initially placed on a regular lattice within each
cell. The first beam's velocities are sampled from the drifting
Maxwellian distribution using the Fortran random number generator
with the fixed seed vector
$s_j=20260810+104729j$, where
$j=1,\ldots,n_{\mathrm{seed}}$ and
$n_{\mathrm{seed}}$ is obtained from \texttt{random\_seed}.
The second beam retains the first beam's particle positions
and reverses all three velocity components, enforcing exact
pairwise symmetry.
The time step is $\Delta t=0.05\,\omega_{pe}^{-1}$,
and the simulation is advanced to $\omega_{pe}t=40$.

An oblique mode is seeded by displacing both beams according to
\begin{equation}
    \boldsymbol{\xi}
    =
    \delta_0\frac{\boldsymbol{k}}{|\boldsymbol{k}|^2}
    \sin(\boldsymbol{k}\cdot\boldsymbol{x}),
    \qquad
    \delta_0=0.005,
\end{equation}
where
$\boldsymbol{k}=2\pi(m_x/L_x,m_y/L_y)$ and
$(m_x,m_y)=(2,1)$.
For a spatially uniform initial distribution, the displacement
produces, to first order,
$\delta n/n_0=-\delta_0\cos(\boldsymbol{k}\cdot\boldsymbol{x})$.
It therefore introduces a controlled $0.5\%$ density perturbation in the
selected mode \cite{birdsall1991}, allowing its growth to be distinguished
from particle noise and compared directly with linear kinetic theory.
Choosing an
oblique mode also exercises deposition, interpolation, and field solution
in both spatial directions rather than reducing the example to an
effectively one-dimensional problem.

\subsection{Component assembly and execution}

\begin{figure}[htbp]
    \centering
    \includegraphics[width=\linewidth]{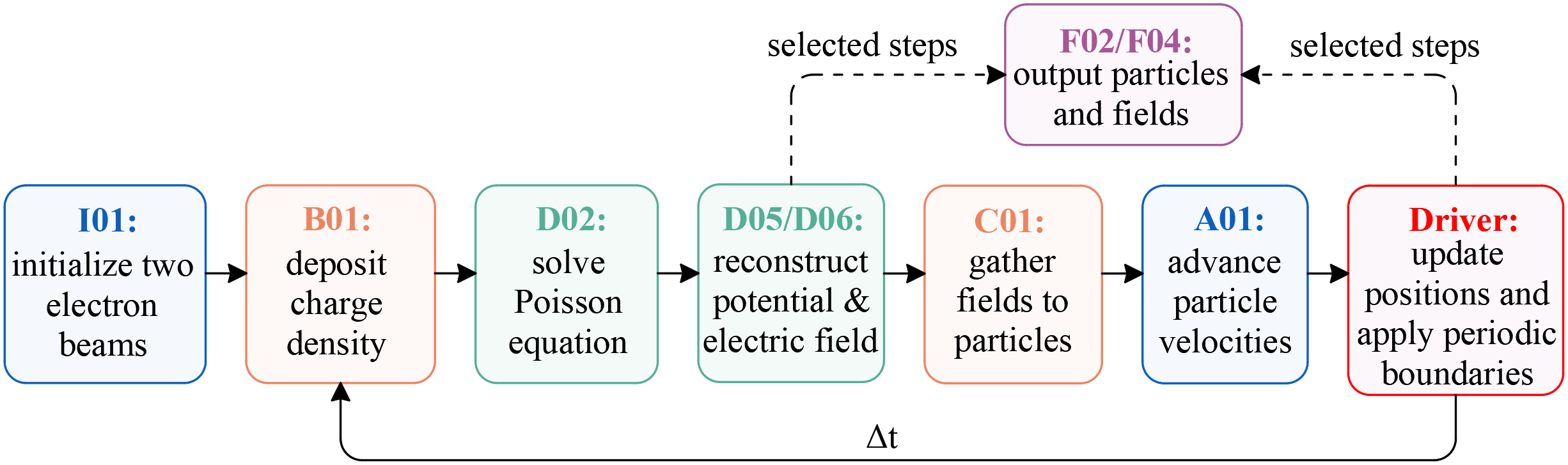}
    \caption{Assembly of AlgoPlasma components in the two-dimensional
    electrostatic PIC example.}
    \label{fig:two_stream_workflow}
\end{figure}

Fig.~\ref{fig:two_stream_workflow} shows the assembly of the numerical
components.
Module \texttt{I01} initializes the particle distribution,
\texttt{B01} deposits the charge density, and \texttt{D02}
solves the Poisson equation.
Modules \texttt{D05} and \texttt{D06} reconstruct the grid
potential and electric field, \texttt{C01} interpolates the
field to the particles, and \texttt{A01} advances their
velocities.
Particle and field data are written by
\texttt{F02}/\texttt{F04}.
The application-specific code manages the data structures,
execution order, particle position updates, periodic boundaries,
and diagnostic frequency.
A single physical cell in $z$ allows this example to reuse the
\texttt{3Dxyz} routines in a 2D3V configuration, in which particle
positions are advanced only in $x$ and $y$ while all three
velocity components are retained.
\ref{app:numerical_method} presents the discrete equations
in dimensional form, together with the solver settings and
boundary conditions.

The example source code is organized into three files:
\path{case_parameters.f90} defines the physical and numerical
parameters, \path{two_stream_case.f90} connects application data
to AlgoPlasma components through stage-level procedures, and
\path{main.f90} specifies the PIC execution sequence.
This separation keeps the workflow explicit and allows individual
numerical operations to be examined or replaced independently.
Together, these files contain 237 physical lines, or 198 source
lines when blank and comment-only lines are excluded.
Listing~\ref{lst:two_stream_driver} reproduces the complete
29-line driver with unnumbered explanatory annotations.

\begin{lstlisting}[
    style=algoplasma,
    caption={Application-level driver of the two-stream example from
\protect\path{examples/001_two_stream_2d/src/main.f90}. Unnumbered
comments identify the underlying AlgoPlasma routines.},
    label={lst:two_stream_driver}
]
program two_stream_2d
    use mpi
    use omp_lib
    use two_stream_parameters
    use two_stream_case
    implicit none
    integer :: step, ierr, nproc

    call mpi_init(ierr)
    call mpi_comm_size(mpi_comm_world,nproc,ierr)
    write(*,'(A,I0,A,I0)') 'MPI ranks: ',nproc,', OpenMP threads: ',omp_get_max_threads()
    call initialize_case (*@\Suppressnumber@*)
    ! I01  sub_I01_par_distribute_equilibrium (*@\Reactivatenumber@*)
    call update_electric_field (*@\Suppressnumber@*)
    ! B01  sub_B01_scatter_3Dxyz
    ! D02  sub_D02_hypre_3Dxyz_bc_A
    !      sub_D02_hypre_3Dxyz_bc_fortran
    ! D05  sub_D05_phi1d_to_phi3d
    ! D06  sub_D06_phi_to_E (*@\Reactivatenumber@*)
    call push_velocities(initial_half_push) (*@\Suppressnumber@*)
    ! C01  sub_C01_gather_3Dxyz
    ! A01  sub_A01_Boris_3Dxyz (*@\Reactivatenumber@*)
    call write_fields(0) (*@\Suppressnumber@*)
    ! F04  sub_F04_field_output_3d_bin (*@\Reactivatenumber@*)
    call write_particles(0) (*@\Suppressnumber@*)
    ! F02  sub_F02_par_output (*@\Reactivatenumber@*)

    do step=1,nt
        call push_velocities(full_push)
        call move_particles(dt) (*@\Suppressnumber@*)
        ! Application-specific position update and periodic boundaries (*@\Reactivatenumber@*)
        call update_electric_field

        if (step==1 .or. mod(step,field_stride)==0) call write_fields(step)
        if (step==1 .or. mod(step,particle_stride)==0) call write_particles(step)
    end do

    call finalize_case (*@\Suppressnumber@*)
    ! D02  sub_D02_hypre_3Dxyz_bc_fortran (finalization) (*@\Reactivatenumber@*)
    call mpi_finalize(ierr)
end program two_stream_2d
\end{lstlisting}

The example runs with one MPI rank and eight OpenMP threads.
MPI supplies the communicator required by HYPRE, while OpenMP
parallelizes the particle loops.
All reported results use 64 bit real arithmetic with GNU Fortran
and a double precision HYPRE build.
The release-specific example documentation records the tested
software and hardware environment, build and run commands,
reference runtime and peak memory usage, and numerical reference
values. It also provides instructions for regenerating
Figs.~\ref{fig:two_stream_phase_space}
and~\ref{fig:two_stream_field_growth}.

\subsection{Results and validation}

\begin{figure}[htbp]
    \centering
    \includegraphics[width=\linewidth]{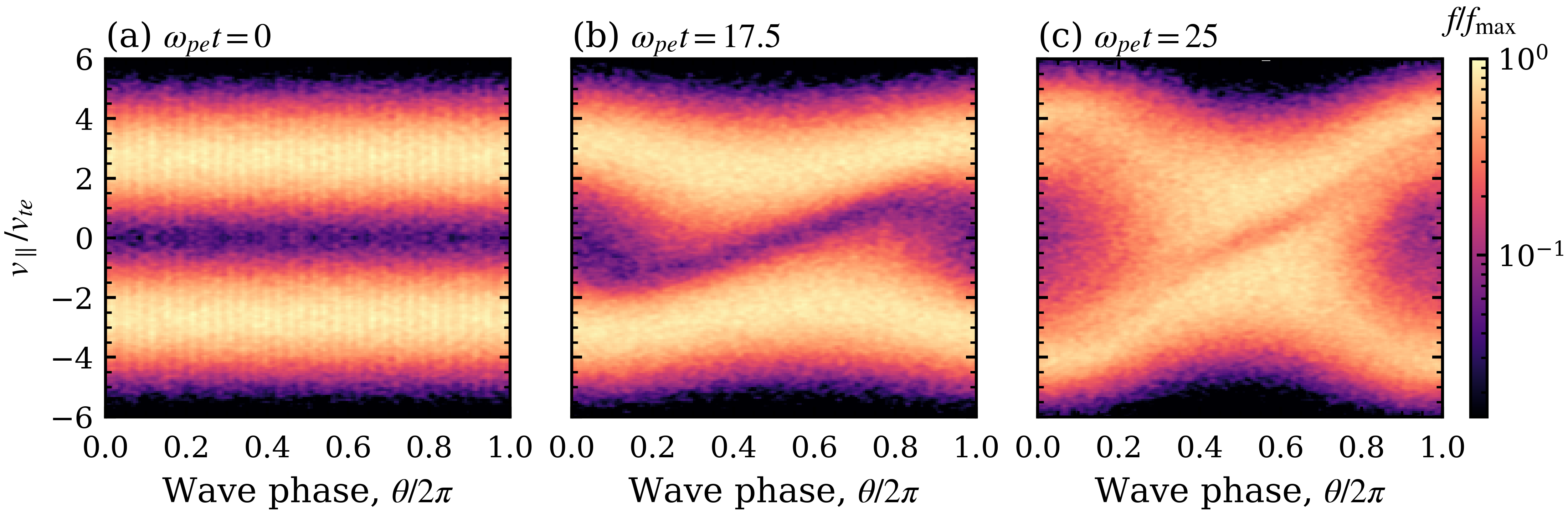}
    \caption{Evolution of the reduced electron distribution in
    wave-aligned phase space at $\omega_{pe}t=0$, $17.5$, and $25.0$.
    The color scale shows Gaussian-smoothed histograms normalized
    by the common maximum over the three panels.}
    \label{fig:two_stream_phase_space}
\end{figure}

The numerical behavior of the assembled application is assessed through
the evolution in phase space, the growth of the seeded electric field mode,
and the global energy balance.
Because the seeded mode is oblique, a conventional $(x,v_x)$
projection mixes particles at different wave phases.
We therefore use the wave-aligned coordinates
\begin{equation}
    \theta
    =
    (\boldsymbol{k}\cdot\boldsymbol{x})\bmod 2\pi,
    \qquad
    v_{\parallel}
    =
    \frac{k_xv_x+k_yv_y}{|\boldsymbol{k}|}.
\end{equation}
These coordinates group particles by wave phase and retain the
velocity component involved in the longitudinal wave--particle
interaction \cite{oneil1965}.
Fig.~\ref{fig:two_stream_phase_space} shows the resulting
evolution in phase space.
The initial beams are centered at
$v_{\parallel}/v_{te}\simeq\pm2.683$.
A coherent, phase-correlated velocity modulation develops by
$\omega_{pe}t=17.5$, followed by strong broadening and
interpenetration at $\omega_{pe}t=25$ as the instability
enters nonlinear saturation
\cite{morse1969warm,morse1969multid}.

\begin{figure}[htbp]
    \centering
    \includegraphics[width=\linewidth]{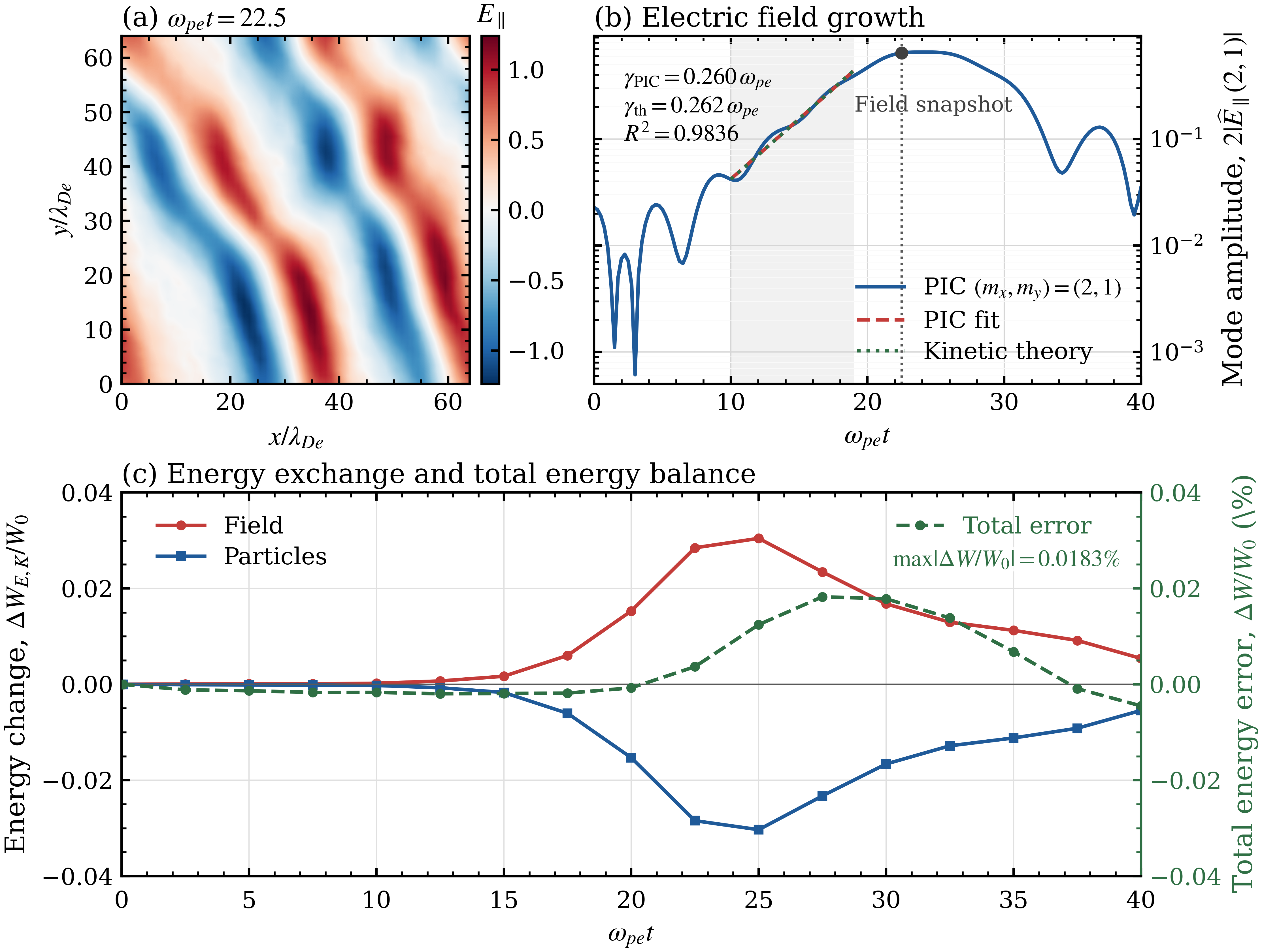}
    \caption{Electric field structure, instability growth, and
    energy balance in the two-stream example.
    (a) Parallel electric field at $\omega_{pe}t=22.5$.
    (b) Evolution of the seeded $(m_x,m_y)=(2,1)$ Fourier amplitude,
    together with the fitted and theoretical growth rates.
    (c) Changes in field and particle kinetic energies (left axis),
    together with the relative total energy error
    (green dashed curve, right axis).}
    \label{fig:two_stream_field_growth}
\end{figure}

The corresponding field diagnostic is the electric field
component along the same wave vector,
\begin{equation}
    E_{\parallel}
    =
    \frac{k_xE_x+k_yE_y}{|\boldsymbol{k}|}.
\end{equation}
Fig.~\ref{fig:two_stream_field_growth}(a) shows this component
at $\omega_{pe}t=22.5$.
Its spatial structure reflects the prescribed oblique
$(2,1)$ mode.

The growth of the seeded mode is measured by
$A_{21}(t)=2|\widehat{E}_{\parallel}(2,1;t)|$, where
\begin{equation}
    \widehat{E}_{\parallel}(m_x,m_y;t)
    =
    \frac{1}{N_xN_y}
    \sum_{i=0}^{N_x-1}
    \sum_{j=0}^{N_y-1}
    E_{\parallel,ij}(t)
    \exp\left[
        -2\pi\mathrm{i}
        \left(
            \frac{m_x\,i}{N_x}+\frac{m_y\,j}{N_y}
        \right)
    \right].
\end{equation}
The factor of two accounts for the conjugate Fourier pair of
the real field.
A least-squares fit of
$\ln A_{21}=a+\gamma_{\mathrm{PIC}}t$
over $10\leq\omega_{pe}t\leq19$, after the initial adjustment
and before nonlinear saturation, gives
\begin{equation}
    \gamma_{\mathrm{PIC}}=0.2603\,\omega_{pe},
    \qquad
    R^2=0.9836.
\end{equation}

For comparison, the warm, symmetric two-stream dispersion
relation \cite{fried1961,morse1969warm} is
\begin{equation}
    \varepsilon(\omega,\boldsymbol{k})
    =
    1+
    \frac{1}{2(k\lambda_{De})^2}
    \sum_{\sigma=\pm1}
    \left[1+\zeta_\sigma Z(\zeta_\sigma)\right]
    =0,
\end{equation}
where
\begin{equation}
    \zeta_\sigma
    =
    \frac{\omega-\sigma k_xu_d}
    {\sqrt{2}\,k v_{te}},
    \qquad
    k=|\boldsymbol{k}|,
\end{equation}
and $u_d=3v_{te}$.
The plasma dispersion function is evaluated as
$Z(\zeta)=\mathrm{i}\sqrt{\pi}\,w(\zeta)$
using the Faddeeva function $w$.
The unstable root is obtained by solving
$\operatorname{Re}\varepsilon=\operatorname{Im}\varepsilon=0$
with a modified Powell hybrid method, yielding
$\gamma_{\mathrm{th}}=0.2616\,\omega_{pe}$.
The PIC growth rate differs from this prediction by less than
$0.5\%$.
Fig.~\ref{fig:two_stream_field_growth}(b) compares the measured
amplitude with the fitted and theoretical exponentials.
For the theoretical curve, the slope is fixed at
$\gamma_{\mathrm{th}}$ and the intercept is fitted over the
same interval.
The mode amplitude peaks near $\omega_{pe}t=24$
and subsequently decreases.

The energy diagnostic provides a complementary check on the
coupled particle--field evolution.
The particle kinetic energy, electrostatic field energy, and
total energy per unit length in $z$ are
\begin{equation}
    W_K
    =
    \frac{1}{2}\sum_p w_p m_e|\boldsymbol{v}_p|^2,
    \qquad
    W_E
    =
    \frac{\varepsilon_0}{2}
    \sum_{i,j}|\boldsymbol{E}_{ij}|^2\Delta x\Delta y,
    \qquad
    W=W_K+W_E,
\end{equation}
where $w_p$ is the particle weight per unit $z$ length
and $\boldsymbol{v}_p$ is the time-centered particle velocity.
Fig.~\ref{fig:two_stream_field_growth}(c) shows the increase
in field energy accompanied by a corresponding decrease in
particle kinetic energy.
All energy changes are normalized by $W_0=W(0)$,
and the relative total energy error is $[W(t)-W_0]/W_0$.
Its maximum recorded absolute value is $0.0183\%$.

For this configuration, the agreement with kinetic theory,
the nonlinear evolution in phase space, and the small energy error
provide a physical check on the PIC workflow assembled from
AlgoPlasma components.

\section{Impact}
\label{impact}

\subsection{Scientific impact}
At present, the clearest scientific impact of AlgoPlasma is its use
in Hall thruster modeling. The numerical routines developed when
it was known as PMSL were used to build PMSL-PIC-HET-3D,
an in-house three-dimensional PIC code introduced and benchmarked
against WarpX in a study of plume-domain effects on azimuthal
instability \cite{chen2025pop}.
Subsequent extensions incorporated realistic magnetic field
configurations \cite{zhao2025magnetic,zhong2026pla}, coupled Monte Carlo
collision and fluid-neutral models \cite{chen2025iepc}, and
self-consistent wall and open-plume treatments
\cite{liu2026nearwall}. Built on this common numerical
foundation, the research progressed from resolving electron drift
instability to quantifying how magnetic topology and boundaries govern
electron transport, constructing effective mobility profiles
\cite{zhao2025transport}, and identifying the field structures responsible
for transport through test-particle and modal analyses
\cite{zhao2026testparticle,zhao2026review}.

Application-driven research has also produced standalone algorithmic
advances. Conservative charge and current deposition on nonuniform
three-dimensional cylindrical meshes was developed together with
continuity and residual self-field diagnostics
\cite{liu2026deposition}. Particle-based and deterministic
free-molecular methods were likewise developed to provide
face-flux-consistent closures for reduced neutral continuity models
\cite{chen2026neutral}. These developments allow numerical artifacts to
be distinguished more clearly from physical behavior and alternative
methods to be compared using common conservation and diagnostic criteria.
They also exemplify the development process underlying AlgoPlasma:
numerical requirements arising in application programs are recast as
separable algorithms with dedicated verification and validation,
providing tested foundations for further research and library expansion.

Within the authors' group, this shared algorithmic base has shifted
software development from repeatedly constructing complete solvers
to reusing established components and adapting the physics,
coupling, and boundary treatments specific to each problem.
This practice reduces duplicate implementation and makes numerical
assumptions easier to examine and compare across application
programs. The complete research application codes cited above
are not included in the public release. The software has not been
commercially deployed or led to the creation of a spin-off company.

\subsection{Maintenance and development}
AlgoPlasma is maintained by the authors' group, with project
maintainers responsible for technical review, integration, and
release decisions. Issue reports are assessed using the reported
version, execution environment, and reproducible examples, with
priority given to numerical correctness and usability.
Contributions are submitted through pull requests and reviewed
for numerical consistency, interface and dependency changes,
documentation, and relevant test results. Accepted changes are
integrated into the \texttt{develop} branch, with tagged releases
prepared at development milestones.

Release notes document interface changes and provide migration
guidance for breaking changes; interfaces are deprecated before
removal where practical. Published versions remain available
for users requiring a fixed release. 
The development roadmap includes broader language support,
graphics processing unit (GPU) acceleration, conservative
charge and current deposition on nonuniform three-dimensional
cylindrical meshes \cite{liu2026deposition}, relativistic
electromagnetic particle--particle methods
\cite{zhang2025empp,zhao2026empp}, and further algorithms for
plasma diagnostic data analysis \cite{peng2026antialiasing}.

\subsection{Current limitations}

Verification and validation coverage varies across components
and remains incomplete. The two-stream example uses one MPI rank
with OpenMP particle loops and does not establish multi-rank
application correctness or parallel scalability. 
Fortran routines using default \texttt{real} declarations rely on
compiler settings for precision, while external interfaces require
compatible data types.
The reported example uses 64 bit real arithmetic with GNU Fortran
on Ubuntu 22.04 LTS; compatibility with other operating systems,
compilers, and dependency configurations is not established
by these results.

Component interfaces may evolve, and API compatibility across
releases is not guaranteed. Documented use remains concentrated
within the authors' group, and meaningful statistics on external
adoption are not yet available.

\section{Conclusions}
\label{conclusion}
AlgoPlasma is an open-source, modular library of reusable numerical
algorithms for plasma modeling.
Its components span particle and field operations,
particle--grid coupling, neutral transport, selected continuity
updates, probe signal analysis, parallel data exchange,
and input/output.
The current release combines these implementations with
documentation and focused verification and validation cases,
supporting selective integration and independent assessment.

The illustrative two-stream example demonstrates
how these components can be assembled into a compact PIC workflow that
reproduces the expected linear growth, nonlinear evolution in phase space,
and energy exchange while remaining close to global energy balance.
Research applications and related algorithmic developments within
the authors' group further demonstrate the versatility of this
algorithm-centered approach across different physical models and
computational settings.
Future releases will incorporate additional
algorithms, physical models, and computational capabilities as they
mature through testing and validation. AlgoPlasma thus provides a
transparent and extensible basis for plasma modeling research,
verification, and education.

\section*{CRediT authorship contribution statement}

Yinjian Zhao: Conceptualization, Methodology, Software, Resources,
Supervision, Project administration, Funding acquisition, Writing --
review \& editing.

Zhongping Zhao: Methodology, Software, Validation, Formal analysis,
Writing -- original draft, Writing -- review \& editing.

Zhe Liu: Methodology, Software, Validation, Formal analysis,
Visualization, Writing -- review \& editing.

Baisheng Wang: Software, Validation, Writing -- review \& editing.

Zilong Peng: Software, Validation, Writing -- review \& editing.

Xin Luo: Software, Validation, Writing -- review \& editing.

Lihuan Xie: Software, Validation, Writing -- review \& editing.

Xi Chen: Methodology, Validation, Investigation, Writing -- review \&
editing.

Zhijun Zhou: Software, Validation, Writing -- review \& editing.

Kunpeng Zhong: Validation, Formal analysis, Investigation, Writing --
review \& editing.

Yingjie Chen: Methodology, Software, Validation, Writing -- review \&
editing.

Changzheng Hu: Methodology, Validation, Writing -- review \& editing.

\section*{Declaration of competing interest}

The authors declare that they have no known competing financial
interests or personal relationships that could have appeared to
influence the work reported in this paper.

\section*{Acknowledgements}

The authors acknowledge support from the National Natural Science
Foundation of China under Grant No. 52472403. This work was also
partially supported by the Harbin Institute of Technology Kunpeng \&
Ascend Center of Cultivation.


\appendix
\section{Component functionality and test availability}
\label{app:component_inventory}

The geometry and dimensionality listed below refer to the
distributed implementations.
For \texttt{K}, the entries describe probe configurations
or inferred wavenumber space rather than spatial simulation meshes.

\begingroup
\scriptsize
\setlength{\tabcolsep}{3pt}
\renewcommand{\arraystretch}{1.15}
\setlength{\LTpre}{0.5\baselineskip}
\setlength{\LTpost}{0.5\baselineskip}

\begin{xltabular}{\textwidth}{
@{}
>{\raggedright\arraybackslash}p{0.15\textwidth}
>{\raggedright\arraybackslash}X
>{\raggedright\arraybackslash}X
>{\raggedright\arraybackslash}p{0.16\textwidth}
>{\centering\arraybackslash}p{0.075\textwidth}
@{}
}

\caption{Functionality and test availability of AlgoPlasma components.}
\label{tab:component_inventory}\\

\toprule
\textbf{Component} &
\textbf{Task / equation or operation} &
\textbf{Numerical method} &
\textbf{Geometry / dimensionality} &
\textbf{Tests} \\
\midrule
\endfirsthead

\multicolumn{5}{@{}l@{}}{
    \tablename~\thetable\ (continued)
}\\
\toprule
\textbf{Component} &
\textbf{Task / equation or operation} &
\textbf{Numerical method} &
\textbf{Geometry / dimensionality} &
\textbf{Tests} \\
\midrule
\endhead

\midrule
\multicolumn{5}{r@{}}{\textit{Continued on next page}}\\
\endfoot

\bottomrule
\endlastfoot

\texttt{A01} &
Nonrelativistic Lorentz force velocity update &
Boris splitting with exact magnetic rotation
\cite{boris1970} &
Cartesian; three velocity components &
Yes \\

\texttt{A02} &
Nonrelativistic particle advancement &
Boris splitting with cylindrical position and basis updates
\cite{boris1970} &
3D cylindrical &
Yes \\

\texttt{A03} &
Relativistic Lorentz force velocity update &
Higuera--Cary pusher \cite{higuera2017} &
Cartesian; three velocity components &
Yes \\

\addlinespace[3pt]

\texttt{B01} &
Particle density and moment deposition &
Linear cloud-in-cell (CIC); cellwise velocity statistics &
Uniform 3D Cartesian &
Yes \\

\texttt{B02} &
Charge and current deposition &
Volume-normalized charge deposition; trajectory-split,
swept-volume current deposition \cite{zhao2023cyl} &
Uniform 3D cylindrical &
Yes \\

\texttt{B03} &
Particle density and weighted moment deposition &
Tensor product B-splines of selectable degree &
Uniform 3D Cartesian &
Yes \\

\addlinespace[3pt]

\texttt{C01} &
Field interpolation and fused particle advancement &
Trilinear CIC interpolation; fused gather--Boris update &
Uniform 3D Cartesian &
Yes \\

\texttt{C02} &
Field interpolation to particles &
Tensor product B-splines of selectable degree &
Uniform 3D Cartesian &
Yes \\

\texttt{C03} &
Interpolation of staggered electric and nodal magnetic fields &
Trilinear interpolation &
Nonuniform 3D cylindrical &
Yes \\

\addlinespace[3pt]

\texttt{D01} &
Poisson solution &
Seven-point finite differences; HYPRE with case-supplied
boundary and solver configuration \cite{falgout2002hypre} &
Uniform 3D Cartesian &
Yes \\

\texttt{D02} &
Poisson solution with boundary corrections &
Seven-point finite differences; HYPRE PFMG
\cite{falgout2002hypre} &
Uniform 3D Cartesian &
Yes \\

\texttt{D03} &
Poisson solution &
Cell-centered finite volumes; HYPRE PFMG
\cite{falgout2002hypre} &
Uniform 3D cylindrical &
Yes \\

\texttt{D04} &
Poisson solution &
Cell-centered finite volumes; HYPRE PFMG
\cite{falgout2002hypre} &
Nonuniform 3D cylindrical &
Yes \\

\texttt{D05} &
Potential array reconstruction and halo exchange &
Vector unpacking; MPI exchange and periodic wrapping &
3D structured subdomains &
No\textsuperscript{a} \\

\texttt{D06} &
Electric field reconstruction,
$\boldsymbol{E}=-\nabla\phi$ &
Second-order centered differences &
Uniform 3D Cartesian &
No\textsuperscript{a} \\

\addlinespace[3pt]

\texttt{E01} &
Maxwell curl updates and absorbing boundaries &
Yee-type FDTD with CPML \cite{yee1966,roden2000cpml} &
Uniform 2D axisymmetric $(r,z)$ &
Yes \\

\texttt{E02} &
Maxwell curl updates and absorbing boundaries &
Yee-type FDTD with CPML \cite{yee1966,roden2000cpml} &
Uniform 3D cylindrical &
Yes \\

\texttt{E03} &
Maxwell curl updates and absorbing boundaries &
Yee FDTD with CPML \cite{yee1966,roden2000cpml} &
Uniform 3D Cartesian &
Yes \\

\addlinespace[3pt]

\texttt{F01} &
Particle data loading &
Per-rank text, binary, and optional HDF5 input &
Six-component particle records &
Yes \\

\texttt{F02} &
Particle data output &
Per-rank text, binary, and optional HDF5 output &
Six-component particle records &
Yes \\

\texttt{F03} &
Field data loading &
Text and binary input &
3D fields; packed or multidimensional arrays &
Yes \\

\texttt{F04} &
Field data output &
Text and binary output; optional node-to-cell averaging &
3D fields; packed or multidimensional arrays &
Yes \\

\addlinespace[3pt]

\texttt{G01} &
Electron--neutral and ion--neutral collisions &
Null-collision Monte Carlo with tabulated cross sections
\cite{vahedi1995} &
3D Cartesian; three velocity components &
Yes\textsuperscript{b} \\

\texttt{G02} &
Collision reaction networks &
Null collision Monte Carlo \cite{vahedi1995};
two- and three-body reactions &
Caller defined cells; three velocity components &
Yes \\

\addlinespace[3pt]

\texttt{H01} &
Field halo exchange &
MPI neighbor exchange with guard value replacement &
3D structured subdomains &
Yes \\

\texttt{H02} &
Particle migration between subdomains &
Two-stage MPI exchange of particle counts and records &
3D structured subdomains &
Yes \\

\texttt{H03} &
Deposited density exchange &
MPI neighbor exchange with boundary value accumulation &
3D structured subdomains &
Yes \\

\addlinespace[3pt]

\texttt{I01} &
Particle position and velocity initialization &
Regular spatial lattice; Box--Muller sampling of
drifting Maxwellians &
Uniform 3D Cartesian &
Yes \\

\texttt{I02} &
Offline particle initialization and subdomain loading &
Density-based rejection sampling; Maxwellian sampling;
binary loading and spatial filtering &
3D Cartesian &
No\textsuperscript{c} \\

\addlinespace[3pt]

\texttt{J01} &
Neutral reference fields and continuity updates &
Free-molecular Monte Carlo;
prescribed-velocity face fluxes;
retained Lax--Friedrichs update &
Reduced $r$--$z$ (2D2V) transport;
2D cylindrical and uniform 3D Cartesian continuity &
Yes \\

\texttt{J02} &
Steady neutral reference fields &
Discrete ordinates;
P1-DG upwind sweeps and wall source iteration &
Reduced $r$--$z$ (2D2V) transport &
Yes \\

\texttt{J03} &
Transient neutral density evolution &
Reference-derived face-flux closure;
upwind finite-volume stepping &
2D $r$--$z$; scalar density &
Yes \\

\addlinespace[3pt]

\texttt{K01} &
Probe spectra and cross-phase statistics &
Segmented FFT; phase statistics &
Paired probe time series &
Yes \\

\texttt{K02} &
Wavenumber--frequency spectrum &
Power-weighted phase histogram &
Single baseline; projected wavenumber &
Yes \\

\texttt{K03} &
Two-dimensional wavevector inference &
Variance-weighted grid MLE using wrapped phase &
Multiple baselines; 2D wavenumber space &
Yes \\

\texttt{K04} &
Breathing waveform and residual &
Record holdout; phase-bin averaging; Fourier synthesis &
Repeated single-point records &
Yes \\

\end{xltabular}

\noindent
\textit{Notes.}
The Tests column records the availability of dedicated test
sources that directly exercise the corresponding component;
it does not indicate complete V\&V coverage.
DG and MLE denote discontinuous Galerkin and maximum likelihood
estimation, respectively.

\smallskip
\noindent
\textsuperscript{a}
\texttt{D05} and \texttt{D06} are exercised in the two-stream
example, but dedicated component tests are not included.

\smallskip
\noindent
\textsuperscript{b}
The \texttt{G01} tests cover cross-section loading and array bounds
handling, not the collision physics.

\smallskip
\noindent
\textsuperscript{c}
The \texttt{I02} checks reproduce binary format and spatial filtering
logic in Python without directly exercising the Fortran routine;
they are therefore not counted as component tests.

\endgroup

\section{Normalization and discrete formulation}
\label{app:numerical_method}

This appendix specifies the normalization and numerical scheme
used in the two-stream example. The discrete equations are written
in dimensional form.

\begin{table}[H]
\centering
\caption{Normalization of the two-stream example.
SI units refer to dimensional quantities; normalized variables
are dimensionless. Particle weights and energies are defined
per unit length in $z$.}
\label{tab:two_stream_normalization}

\begingroup
\scriptsize
\setlength{\tabcolsep}{4pt}
\renewcommand{\arraystretch}{1.15}

\begin{tabular*}{\linewidth}{
@{\extracolsep{\fill}}llll@{}
}
\toprule
\textbf{Quantity} &
\textbf{Normalized variable} &
\textbf{Reference scale} &
\textbf{SI unit} \\
\midrule

Position &
$\boldsymbol{x}/\lambda_{De}$ &
$\lambda_{De}$ &
m \\

Time &
$\omega_{pe}t$ &
$\omega_{pe}^{-1}$ &
s \\

Velocity &
$\boldsymbol{v}/v_{te}$ &
$v_{te}$ &
m\,s$^{-1}$ \\

Number density &
$n/n_0$ &
$n_0$ &
m$^{-3}$ \\

Charge density &
$\rho/(en_0)$ &
$en_0$ &
C\,m$^{-3}$ \\

Potential &
$e\phi/(k_{\mathrm B}T_e)$ &
$k_{\mathrm B}T_e/e$ &
V \\

Electric field &
$\boldsymbol{E}/E_{\mathrm{ref}}$ &
$E_{\mathrm{ref}}=k_{\mathrm B}T_e/(e\lambda_{De})$ &
V\,m$^{-1}$ \\

Wavenumber &
$\boldsymbol{k}\lambda_{De}$ &
$\lambda_{De}^{-1}$ &
m$^{-1}$ \\

Frequency; growth rate &
$\omega/\omega_{pe}$; $\gamma/\omega_{pe}$ &
$\omega_{pe}$ &
s$^{-1}$ \\

Particle weight &
$w_p/(n_0\lambda_{De}^2)$ &
$n_0\lambda_{De}^2$ &
m$^{-1}$ \\

Energy per unit $z$ length &
$W/(n_0k_{\mathrm B}T_e\lambda_{De}^2)$ &
$n_0k_{\mathrm B}T_e\lambda_{De}^2$ &
J\,m$^{-1}$ \\

\bottomrule
\end{tabular*}

\endgroup
\end{table}

\begin{itemize}
\setlength{\itemsep}{4pt}

\item \textit{Particle--grid coupling.}
Charge deposition and field interpolation use the same linear
cloud-in-cell (CIC) weights:
\begin{equation}
    S(\xi)=\max(0,1-|\xi|),
    \qquad
    W_{ij,p}
    =
    S\!\left(\frac{x_p-x_i}{h}\right)
    S\!\left(\frac{y_p-y_j}{h}\right),
\end{equation}
\begin{equation}
    \rho_{ij}
    =
    en_0-\frac{e}{h^2}\sum_p w_pW_{ij,p},
    \qquad
    \boldsymbol{E}_p
    =
    \sum_{i,j}W_{ij,p}\boldsymbol{E}_{ij}.
\end{equation}
Here, $h=\Delta x=\Delta y=\lambda_{De}$,
$(x_i,y_j)=((i-\tfrac12)h,(j-\tfrac12)h)$ are cell centers,
and $w_p=n_0L_xL_y/N_p$ is the uniform particle weight per unit
$z$ length, with $N_p$ the total particle count.
The weights are extended periodically in $x$ and $y$.

\item \textit{Poisson solution.}
The second-order five-point discretization is
\begin{equation}
    \frac{
    4\phi_{ij}-\phi_{i-1,j}-\phi_{i+1,j}
    -\phi_{i,j-1}-\phi_{i,j+1}
    }{h^2}
    =
    \frac{\rho_{ij}-\langle\rho\rangle}{\varepsilon_0},
    \qquad
    \langle\phi\rangle=0,
\end{equation}
where $\langle\cdot\rangle$ denotes the grid average.
The right-hand side is demeaned before solution, and the potential
afterward, enforcing periodic solvability and a zero-mean gauge.
HYPRE's PFMG solver uses a relative residual tolerance of $10^{-10}$.
The single physical cell in $z$ has homogeneous Neumann conditions
on both faces, reducing the stencil to two dimensions.

\item \textit{Electric field reconstruction.}
Centered differences give
\begin{equation}
    E_{x,ij}
    =
    -\frac{\phi_{i+1,j}-\phi_{i-1,j}}{2h},
    \qquad
    E_{y,ij}
    =
    -\frac{\phi_{i,j+1}-\phi_{i,j-1}}{2h},
\end{equation}
with $E_z=0$.

\item \textit{Particle advancement.}
Leapfrog staggering places positions and fields at integer time
levels and velocities at half-integer levels.
For electrons with $q_e=-e$ and $\boldsymbol{B}=0$,
\begin{equation}
\begin{aligned}
    \boldsymbol{v}_p^{n+1/2}
    &=
    \boldsymbol{v}_p^{n-1/2}
    -\frac{e\Delta t}{m_e}\boldsymbol{E}_p^n,\\
    \boldsymbol{x}_p^{n+1}
    &=
    \boldsymbol{x}_p^n
    +\Delta t\,\boldsymbol{v}_{p,xy}^{n+1/2}.
\end{aligned}
\end{equation}
Here, $\boldsymbol{x}_p=(x_p,y_p)$ and
$\boldsymbol{v}_{p,xy}=(v_{x,p},v_{y,p})$ are the in-plane
position and velocity.
Initialization applies
$\boldsymbol{v}_p^{-1/2}
=\boldsymbol{v}_p^0-q_e\Delta t\,\boldsymbol{E}_p^0/(2m_e)$.
Particle positions are wrapped modulo $L_x$ and $L_y$,
with $z_p=h/2$ fixed.

\end{itemize}

\bibliographystyle{elsarticle-num}
\bibliography{references}




\end{document}